\documentclass[11pt]{article}

\usepackage{geometry}  
\usepackage[T1]{fontenc}  
\usepackage[utf8x]{inputenc}  
\usepackage{lineno}  
\usepackage{fancyhdr}  
\usepackage{enumitem}  
\usepackage{hyperref}  
\usepackage{titling}  
\usepackage[sort&compress]{natbib}  
\usepackage{mathtools}  
\usepackage{titlesec}  
\usepackage{lastpage}  

\usepackage[english]{babel}  
\usepackage{amsmath}  
\usepackage{amsfonts}  
\usepackage{amssymb}  
\usepackage{wasysym}  
\usepackage{bbm}  
\usepackage{array}  
\usepackage{xr}  
\usepackage{verbatim} 
\usepackage{float} 

\usepackage[table]{xcolor}
\usepackage{array}
\newcommand{\PreserveBackslash}[1]{\let\temp=\\#1\let\\=\temp}
\newcolumntype{C}[1]{>{\PreserveBackslash\centering}p{#1}}
\newcolumntype{R}[1]{>{\PreserveBackslash\raggedleft}p{#1}}
\newcolumntype{L}[1]{>{\PreserveBackslash\raggedright}p{#1}}
\usepackage{subfigure}

\hypersetup{%
  pdfborderstyle={/S/U/W 1}
}

\setcitestyle{square,numbers}
\titleformat{name=\section}{\normalfont\Large\bfseries}{}{0pt}{}
\titleformat{name=\subsection}{\normalfont\large\bfseries}{}{0pt}{}
\titleformat{name=\subsubsection}{\normalfont\normalsize\bfseries}{}{0pt}{}

\newcommand{\email}[1]{\href{mailto:{#1}}{{#1}}}

\newcommand{\keywords}[1]{\textbf{Keywords}: {#1}}

\newcommand{\wordcount}[2]{\begin{tabular}{rl}%
\textbf{Manuscript word count}: 	& {#1}\\
\textbf{Abstract word count}: 		& {#2}\\
\end{tabular}}

 \newcommand{\optincludegraphics}[2][{}]{\includegraphics[{#1}]{{#2}}}

\newcommand{\capt}[2][]{\caption[{#1}]{\textbf{{#1}}\newline{#2}}}

\newtagform{brackets}{[}{]}
\usetagform{brackets}

\newcommand{\thejournal}[1]{Magnetic Resonance in Medicine}

\title{Build-A-FLAIR: Synthetic T$_2$-FLAIR Contrast Generation through Physics Informed Deep Learning}

\renewcommand{\jobname}{BuildAFlair_MRI}
\immediate\write18{texcount -sum=1,1,1,1,1,1,1 -merge -incbib -dir -utf8 \jobname.tex > \jobname.wcTotal }
\immediate\write18{texcount -sub=none -merge -incbib -dir -utf8 \jobname.tex | grep _main_ | grep -oE '[0-9]+' | python -c "import sys; print(sum(int(l) * w for l, w in zip(sys.stdin, [1, 1, 0, 0, 0, 0, 0])))" > \jobname.wcManuscript }
\immediate\write18{texcount -sub=none -merge -incbib -dir -utf8 \jobname.tex | grep Abstract | grep -oE '[0-9]+' | python -c "import sys; print(sum(int(l) * w for l, w in zip(sys.stdin, [1, 1, 0, 0, 0, 0, 0])))" > \jobname.wcAbstract }

\begin{document}

\begin{titlepage}
{\noindent\LARGE\bf \thetitle}

\bigskip

\begin{flushleft}\large
	Andrew S. Nencka, PhD. \textsuperscript{1,2,{*}},
	Andrew Klein, MD \textsuperscript{1},
	Kevin M. Koch, PhD. \textsuperscript{1,2},
	Sean D. McGarry \textsuperscript{1}, 
	Peter S. LaViolette, PhD. \textsuperscript{1},
	Eric S. Paulson, PhD. \textsuperscript{3},
	Nikolai J. Mickevicius, PhD. \textsuperscript{3},
	L. Tugan Muftuler, PhD. \textsuperscript{4,2},
	Brad Swearingen \textsuperscript{4},
	Michael McCrea, PhD.\textsuperscript{4}
\end{flushleft}

\bigskip

\noindent
\begin{enumerate}[label=\textbf{\arabic*}]
\item Department of Radiology, Medical College of Wisconsin, Milwaukee, WI, USA
\item Center for Imaging Research, Medical College of Wisconsin, Milwaukee, WI, USA
\item Department of Radiation Oncology, Medical College of Wisconsin, Milwaukee, WI, USA
\item Department of Neurosurgery, Medical College of Wisconsin, Milwaukee, WI, USA
\end{enumerate}

\bigskip


\textbf{*} Corresponding author:

\indent\indent
\begin{tabular}{>{\bfseries}rl}
Name		& Andrew S. Nencka	\\
Department	& Department of Radiology \\
	        & Center for Imaging Research \\
            & Medical College of Wisconsin \\
Address 	& 8701 Watertown Plank Road	\\
			& Milwaukee WI 53226	\\
            & USA	\\
E-mail		& \email{anencka@mcw.edu}\\
\end{tabular}

\vfill

\wordcount{2795}{200}


\end{titlepage}

\pagebreak

\begin{abstract}



\textbf{Purpose}: Magnetic resonance imaging (MRI) exams include multiple series with varying contrast and redundant information. For instance, T$_2$-FLAIR contrast is based upon tissue T$_2$ decay and the presence of water, also present in  T$_2$- and diffusion-weighted contrasts. T$_2$-FLAIR contrast can be hypothetically modeled through deep learning models trained with diffusion- and T$_2$-weighted acquisitions.

\textbf{Methods}: Diffusion-, T$_2$-, T$_2$-FLAIR-, and T$_1$-weighted brain images were acquired in 15 individuals. A convolutional neural network was developed to generate a T$_2$-FLAIR image from other contrasts. Two datasets were withheld from training for validation.

\textbf{Results}: Inputs with physical relationships to T$_2$-FLAIR contrast most significantly impacted performance. The best model yielded results similar to acquired T$_2$-FLAIR images, with a structural similarity index of 0.909, and reproduced pathology  excluded from training.  Synthetic images qualitatively exhibited lower noise and increased smoothness compared to acquired images. 

\textbf{Conclusion}: This suggests that with optimal inputs, deep learning based contrast generation performs well with creating synthetic T$_2$-FLAIR images. Feature engineering on neural network inputs, based upon the physical basis of contrast, impacts the generation of synthetic contrast images. A larger, prospective clinical study is needed.

\end{abstract}

\bigskip

\pagebreak

\section{Introduction}

A benefit of magnetic resonance imaging (MRI) is the availability of varying contrasts based upon physical properties of the tissue being imaged \citep{bernstein}. Thus, MRI exams include the acquisition of several different imaging series with varying contrasts \citep{bushberg}. With each series requiring multiple minutes of acquisition time, this leads to long exam times. Long exam times yield increased imaging costs, decreased patient tolerance, and decreased access to MRI. Much work is being done to reduce MRI exam duration to address these challenges, thereby improving the value of MRI. 

One way to reduce exam duration is to shorten each acquired series. Series acquisitions can be accelerated by acquiring less data, either reducing anatomical coverage \citep{ivi, zoom, zoom2, focus} or spatial resolution. Because of Fourier encoding used in MRI, there is a Nyquist sampling criterion which must be met to enable the reconstruction of artifact free images \citep{bernstein}.  Parallel imaging, utilizing coil array sensitivities to spatially encode the image, allows model-based reconstruction of images sampled below the Nyquist limit \citep{sense, grappa, sms}. Sampling beyond the Nyquist limit can additionally be achieved using the mathematics of image compression and iterative reconstruction algorithms through compressed sensing reconstruction \citep{cs}. Beyond compressed sensing reconstruction, empirical methods using artificial intelligence and machine learning are also now pushing the boundaries of parallel imaging forward \citep{mlRecon}.

Exam duration can be further reduced by decreasing the number of acquired series.  Multiple thick slice 2D acquisitions with matching contrast and orthogonal planes of acquisition are being replaced by single 3D acquisitions that can be reformatted in arbitrary planes and include higher signal-to-noise ratio to enable further use of parallel imaging \cite{2dTo3d_1}. Exam duration can also be reduced by including the reconstruction of multiple contrasts from specialized acquisitions, like synthetic MRI \citep{synthMRI1, synthMRI2} and MR fingerprinting \citep{fingerprinting}. 

This work aims to improve MRI value by reducing the number of series acquired in an MRI exam through the synthetic generation of  contrasts from other series acquired in the standard of care. The complexity of the underlying biology and physics makes the mapping of a set of MRI contrast weighted images to a different contrast weighted image a difficult analytical problem. In such a case, a complicated non-linear transform could be highly unstable and highly dependent upon system settings including various transmit and receive gains which vary from patient-to-patient. Instead of developing such an analytical solution, we present a deep convolutional neural network which generates an image with new contrast based upon input images of other contrasts.

This study tests two hypotheses. First, we determine whether a convolutional neural network can be developed  to yield accurate T2-FLAIR images from other contrasts. Second, we test the impact of input contrast selection on neural network performance, hypothesizing that inclusion of contrasts arising from similar physical properties as the desired output improves model performance.




\section{Methods}


The data used for this proof of concept implementation were acquired as part of a large-scale study of sports related concussion \citep{h2h2_1, h2h2_2, h2h2_3}. This source was selected because it includes many high-resolution 3D image acquisitions across a number of imaging contrasts with well controlled acquisition parameters. A subset of the data acquired in this protocol, including sessions from 15 male collegiate athletes acquired in the fall of 2017 on a GE Healthcare Discovery MR750 running software release DV26R01, were extracted for this study.

The one hour exam in the larger study included T$_1$-weighted MPRAGE, T$_2$-weighted 3D variable flip angle fast spin echo (Cube), T$_2$-FLAIR Cube, multi-echo susceptibility weighted (SWAN), diffusion tensor, arterial spin labeling, and task-free fMRI. Here, an MPRAGE (184 sagittal 1mm$^3$ isotropic resolution , TE 2ms, TR 4700ms, TI 1060ms), T2-weighted Cube (180 sagittal 1mm$^3$ isotropic resolution, TE 93ms, TR 2505ms, ETL 140), T$_2$ FLAIR Cube (180 sagittal 1mm$^3$ isotropic resolution, TE 118ms, TR 6002ms, TI 1600ms, ETL 230), and diffusion tensor imaging (DTI; 47 sagittal 3mm$^3$ isotropic resolution, TE 67ms, TR 5250ms, 30 directions b=1000mm$^2$/s, 30 directions b=2000mm$^2$/s) series were analyzed. 

Images were converted from DICOM to NIFTI format using dcm2niix \cite{dcm2nii}. Diffusion images were processed using a standard pipeline in FSL \citep{fsl} to yield fractional anisotropy (FA), mean diffusivity (MD), and baseline T$_2$ weighted images (S0). Images within each subject were registered to the MPRAGE volume using FLIRT \citep{flirt}. Registered NIFTI images were read into Python 3.6 \citep{python} as NumPy arrays \citep{numpy} using the NIBabel package \citep{nibabel}. Images derived from the diffusion acquisition were resampled with third order splines to 1mm$^3$ resolution using functions in NIBabel.

\begin{figure}[H] 
\optincludegraphics[width=\textwidth]{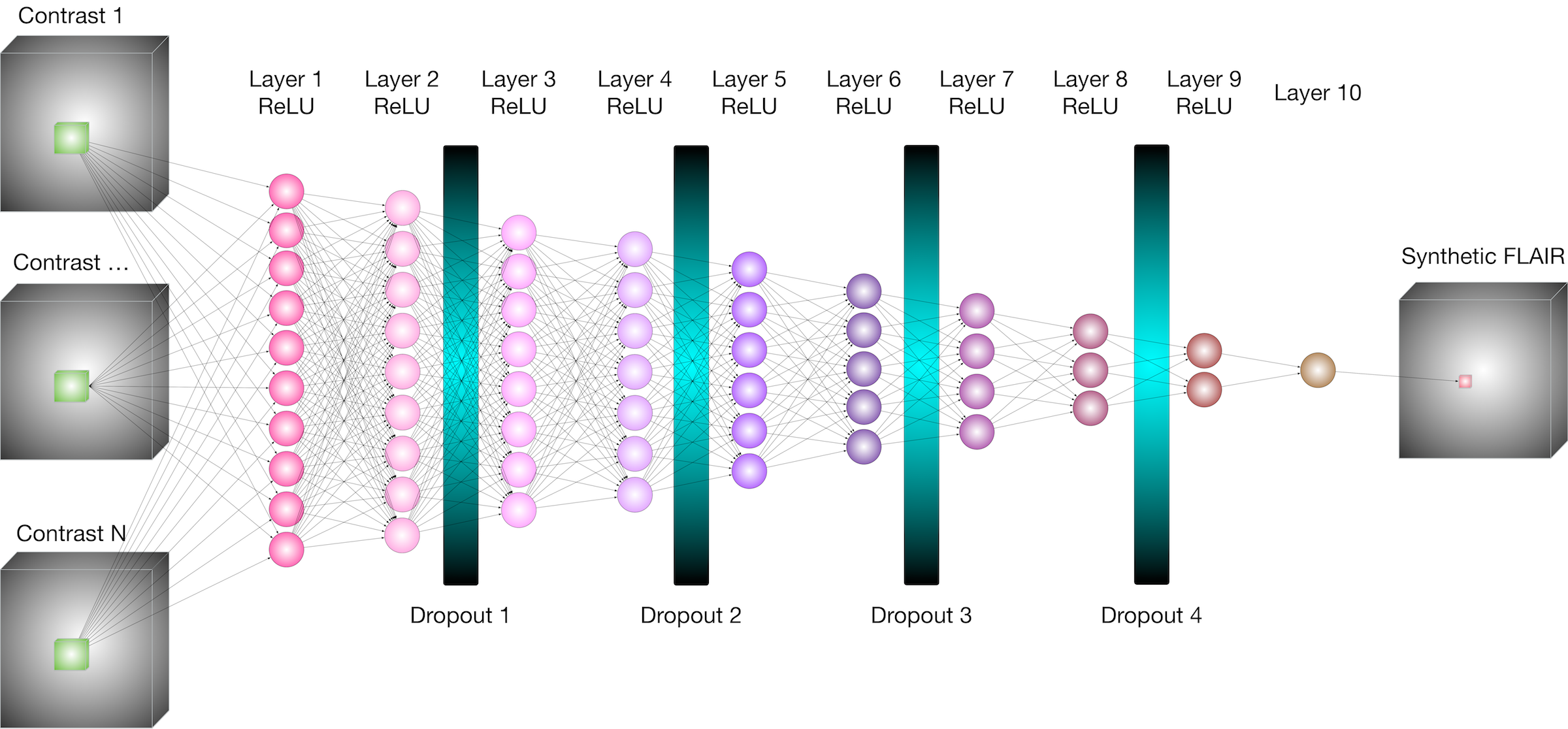}
\capt[The Build-A-FLAIR Network.]{The Build-A-FLAIR network ingests local 3D volumes of multiple contrast images and outputs a single voxel value. This convolutional network estimates each output voxel independently, and consists of 10 dense layers with ReLU activations and 4 dropout layers to prevent overfitting.}
\label{fig:network}
\end{figure}

The Build-A-FLAIR deep neural network was developed using PyTorch \citep{pytorch}. The model was a patch-to-voxel convolutional neural network, modeling each output voxel value through a dense neural network with inputs of 3-dimensional neighborhoods ($5 \times 5 \times 5$ voxels) about the voxel from the several input contrast images. The network includes ten densely connected layers, linearly decreasing in number of neurons from the number of input voxels to one across the network. Each artificial neuron was activated with a rectified linear unit (ReLU, \citep{relu}). Dropout layers with 50\% dropout were added after the second, fourth, sixth, and eighth layers during training to reduce over-fitting during training \citep{dropout}. A graphical representation of this Build-A-FLAIR network is shown in Figure \ref{fig:network}.

\begin{figure}[H] 
\mbox{
    \subfigure[T$_2$-Weighted]{
            \optincludegraphics[width=0.3\textwidth]{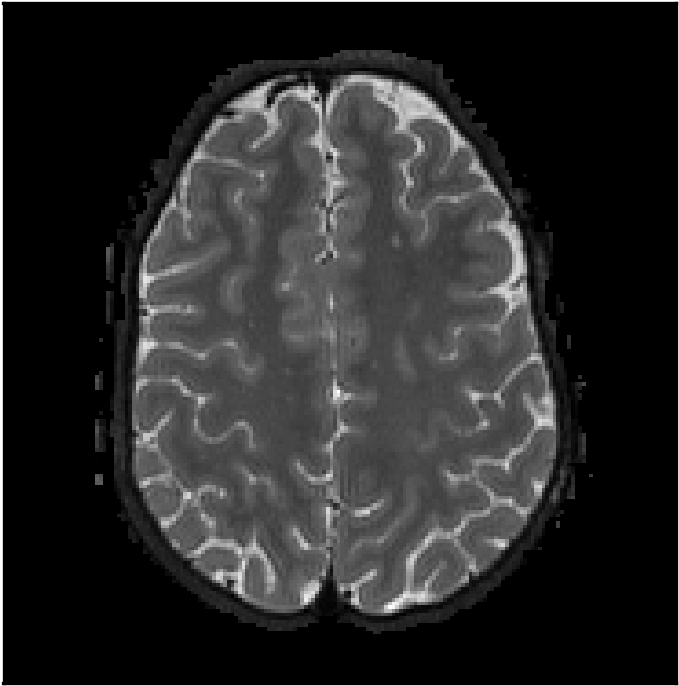}}
    \subfigure[T$_1$-Weighted]{
        \optincludegraphics[width=0.3\textwidth]{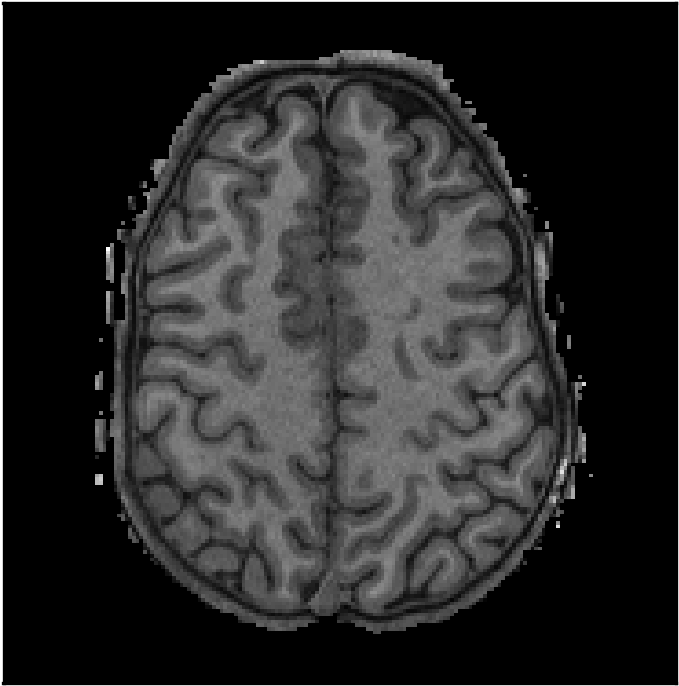}}
    \subfigure[T$_2$-FLAIR]{
        \optincludegraphics[width=0.3\textwidth]{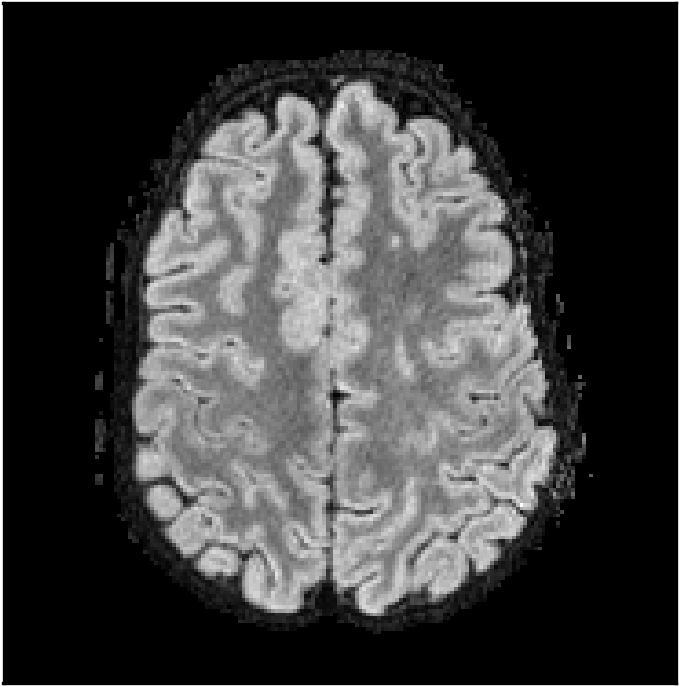}}
    } \\ 
    \mbox{
    \subfigure[MD]{
            \optincludegraphics[width=0.3\textwidth]{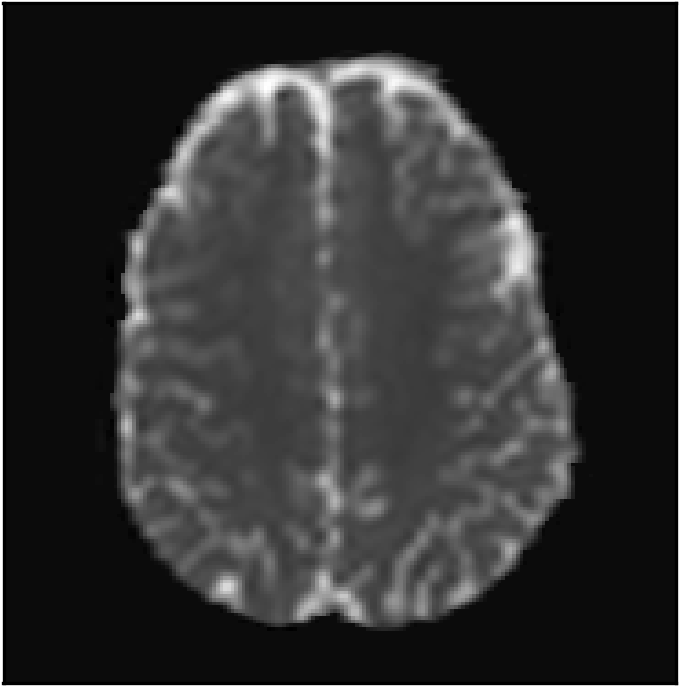}}
    \subfigure[FA]{
        \optincludegraphics[width=0.3\textwidth]{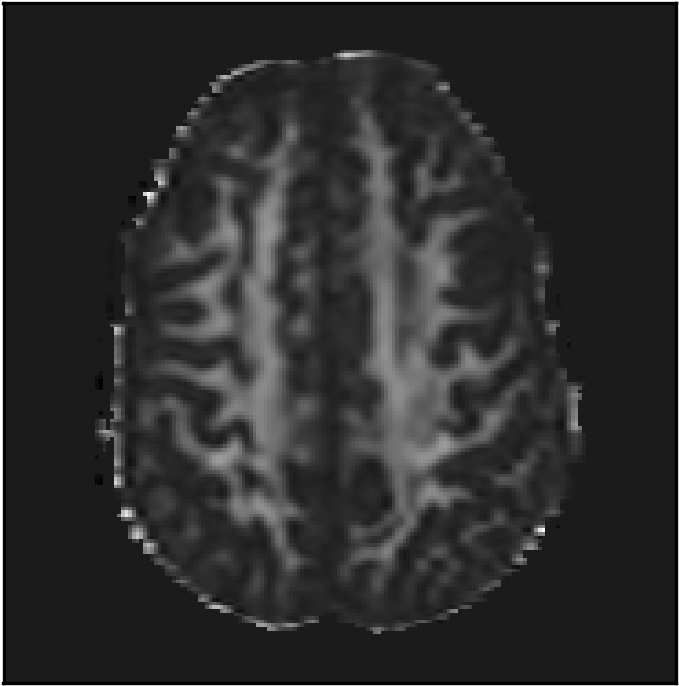}}
    \subfigure[S0]{
        \optincludegraphics[width=0.3\textwidth]{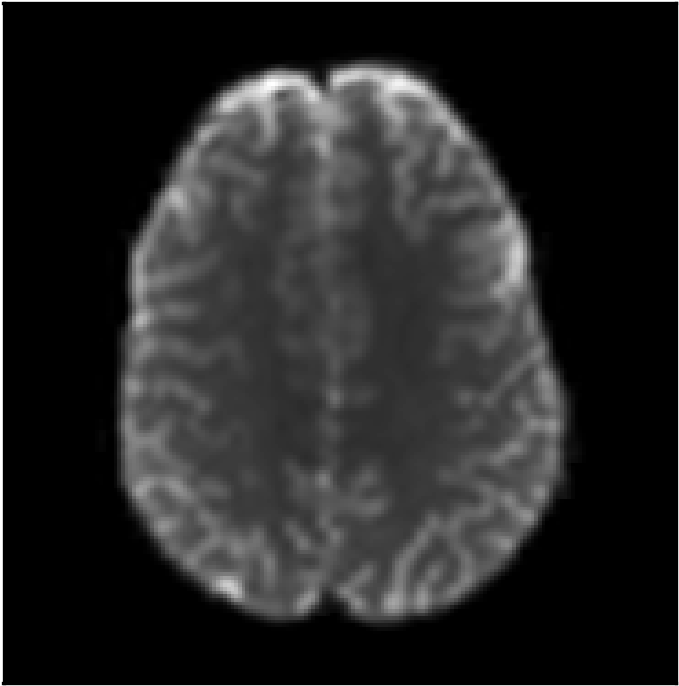}}
    } 
\capt[Data from a validation subject (V1) input to the Build-A-FLAIR network.]{A T$_2$-FLAIR hyperintensity is apparent in the participant's left (image right) frontal lobe. This dataset was not included the the training phase of model development.}
\label{fig:ideal}
\end{figure}

Data for thirteen participants were used for network training. One participant, who exhibited an asymptomatic white matter hyperintensity in the frontal lobe, was excluded from the training process and reserved for algorithm validation (referred to as V1 below). Images from this participant are shown in Figure \ref{fig:ideal}, with the white matter hyperintensity shown in the participant's left anterior frontal lobe. Because a key feature of synthetic contrast generation is the reproduction of pathology, this particular subject was selected for algorithm validation. The other 14 subjects were free from observed pathology. A random subject was also excluded from the training dataset for validation to reduce the probability of over-fitting (referred to as V2 below).

Training was performed on a computer with an Intel i7-3770K processor and an NVIDIA Titan V graphical processing unit. Training included 500 epochs and batch sizes of 100 estimated voxels. For each epoch, images from one subject of the training set were randomly selected and 2,000 voxels to estimate were randomly extracted without resampling from within the brain from that subject. An Adam optimizer was used with a learning rate of 1e-4 \citep{adam}, and the mean squared error of the estimated voxel values was minimized. Following each epoch, 2,000 voxels from the validation dataset V2 were estimated and the model was saved if the mean squared error on the validation dataset was reduced.

Ten models were tested. These models were designed to evaluate the hypothesis that performance is optimized with input contrasts physically related to the output contrast. The models are described in Table \ref{tab:models}. While the network architecture and training procedure were controlled from model-to-model, the number of inputs, and thus number of neurons in each layer, varied with each model. For each input contrast, 125 voxels corresponding to the $5 \times 5 \times 5$ voxel neighborhood around the voxel to be generated were included in the input. Thus, the first layer  included 125 neurons for a network including only the T$_2$-weighted acquisition for an input while that layer included 250 neurons for a model including both T$_2$-weighted and T$_1$-weighted inputs.

Following training, fit models were applied to the V1 validation dataset to generate a full synthetic T$_2$-FLAIR volume. The structural similarity index (SSIM) \citep{ssim} was calculated over the brain between the synthetic and acquired T$_2$-FLAIR image volumes.

\section{Results}


Performances of the models are shown in Table \ref{tab:models}. Models with single contrast inputs (T$_2$-, $T_1$-, or diffusion-weighted) were among the worst performing models, with the three lowest SSIM metrics. Models built upon only T$_2$-weighted and T$_1$ weighted anatomical acquisitions performed nearly as poorly as the model built with only diffusion data as an input. 

Models including both high resolution anatomical images and diffusion imaging metrics yielded superior performance. If only one high resolution anatomical imaging dataset was included in the model, performance was better if it was T$_2$-weighted. With diffusion metrics included, models with inputs of both T$_1$-weighted and T$_2$-weighted images yielded marginally improved results over a model without $T_1$-weighted images. In all cases, including FA yielded poorer performance compared to equivalent models without FA.

\begin{figure}[H] 
\mbox{
    \subfigure[Acquired Axial]{
            \optincludegraphics[width=0.3\textwidth]{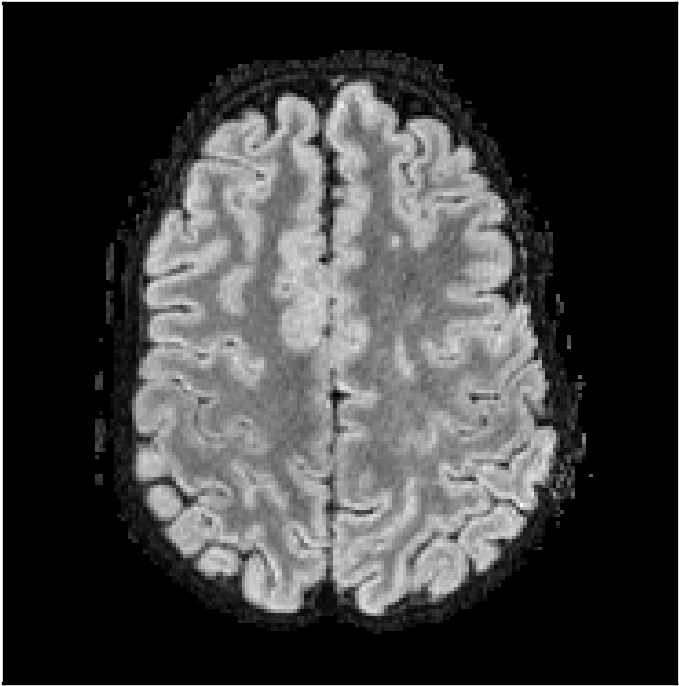}}
    \subfigure[Acquired Coronal]{
        \optincludegraphics[width=0.3\textwidth]{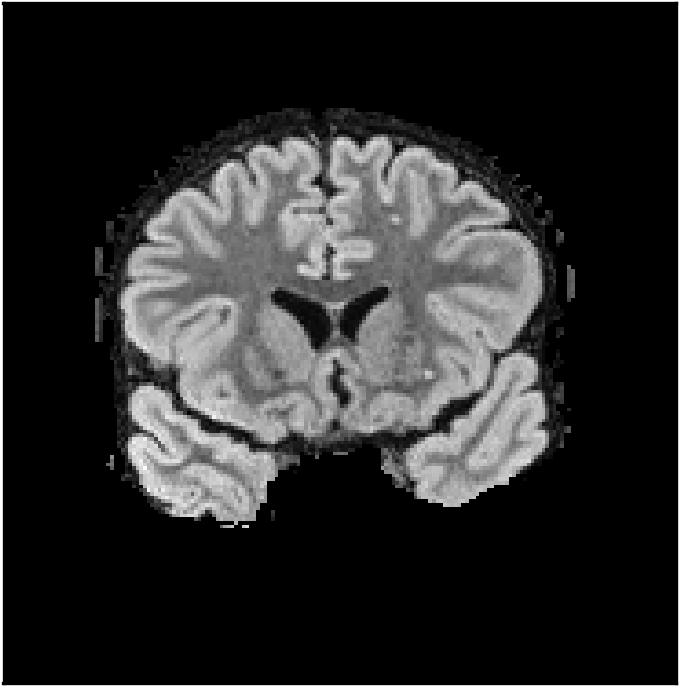}}
    \subfigure[Acquired Sagittal]{
        \optincludegraphics[width=0.3\textwidth]{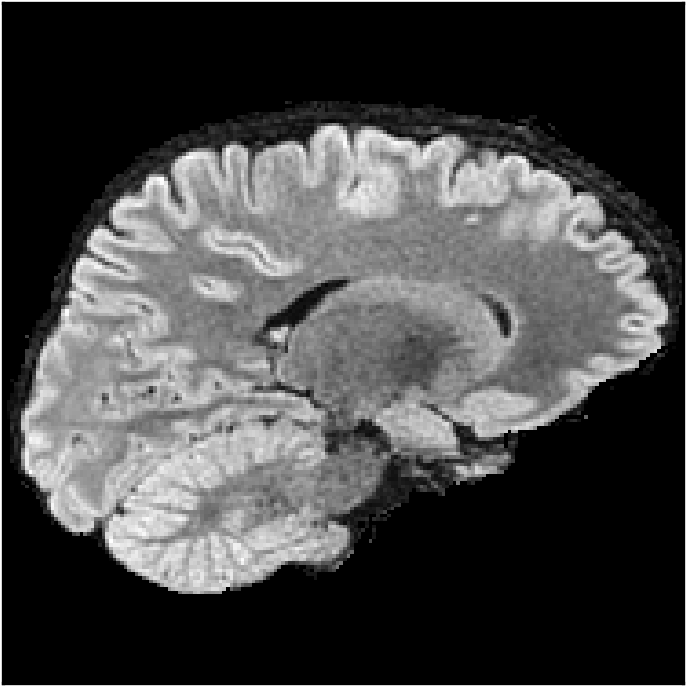}}
    } \\ 
    \mbox{
    \subfigure[Synthetic Axial]{
            \optincludegraphics[width=0.3\textwidth]{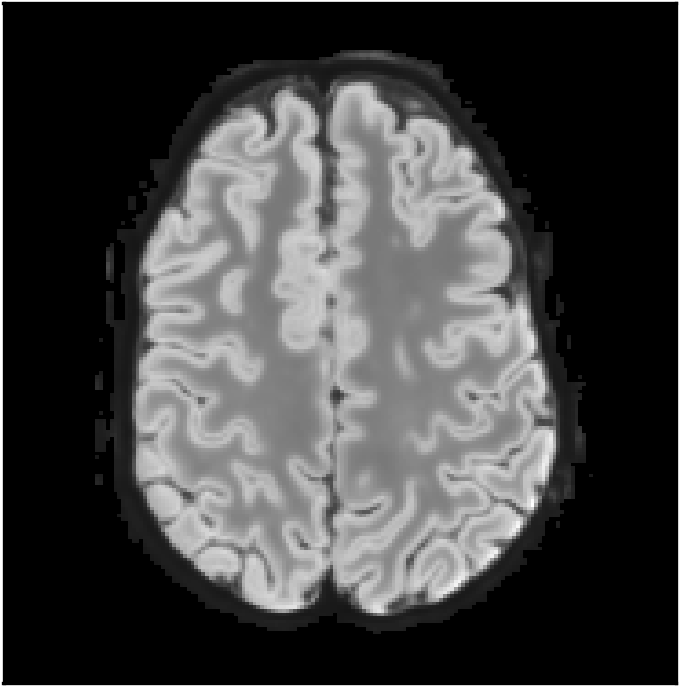}}
    \subfigure[Synthetic Coronal]{
        \optincludegraphics[width=0.3\textwidth]{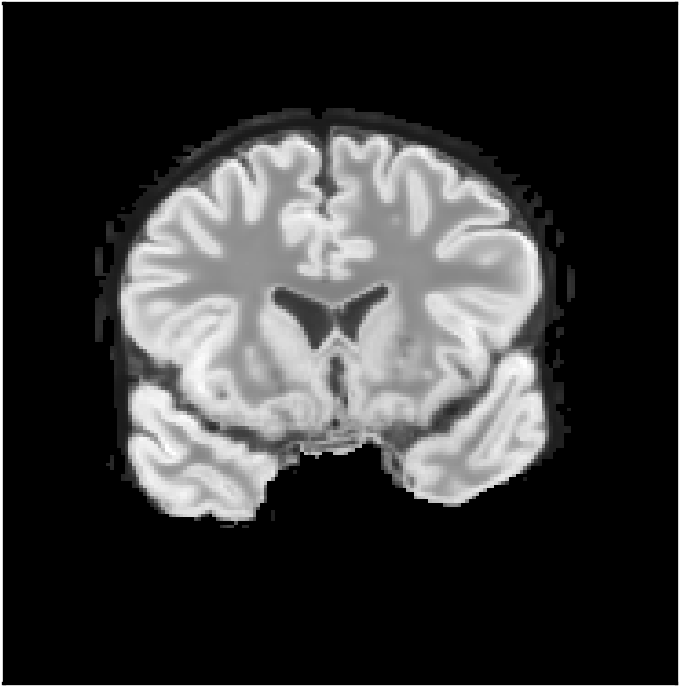}}
    \subfigure[Synthetic Sagittal]{
        \optincludegraphics[width=0.3\textwidth]{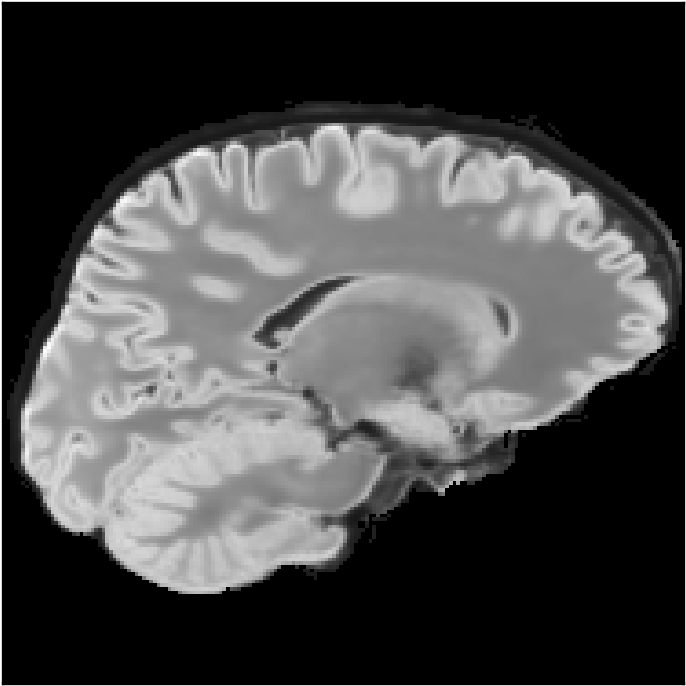}}
    } 
\capt[Build-A-FLAIR output and ideal images from the validation subject (V1).]{This validation subject was segregated from all training procedures, in part due to the presence of a asymptomatic white matter hyperintensity. The white matter hyperintensity is reproduced in the Build-A-FLAIR output.}
\label{fig:valid}
\end{figure}

Synthetic FLAIR images resulting from the best performing Build-A-FLAIR network applied to validation subject V1 are shown in Figure \ref{fig:valid}, with acquired images in panels (a-c) and synthetic images shown in Figure panels (d-f). The aforementioned hyperintensity is visible on all cross sections in the acquired images, and was reproduced in the synthetic T$_2$-FLAIR image derived from the Build-A-FLAIR network. The synthetic images qualitatively exhibit more smoothness and less noise than the acquired T$_2$-FLAIR images.

\section{Discussion}


This study demonstrates the utility of a convolutional neural network for generating synthetic T$_2$-FLAIR images from conventionally acquired images of different contrast. The model reproduces pathology not present in training data. Importantly, optimal network performance results from the inclusion of physically relevant contrasts relating to T$_2$-FLAIR contrast. In fact, inclusion of images in the training that were not directly physically relevant to  T$_2$-FLAIR contrast were detrimental to model performance. While models employing deep learning are widely considered to be ``black boxes,'' these results shed some light into the underlying mechanisms of such models.

Using deep learning to transform image contrast is not new. In the field of artificial intelligence, there has been great progress in developing techniques for \emph{style transfers} \citep{styleTransfer}. With style transfers, the characteristics of the desired output \emph{style} are learned from an image with the desired output style and the characteristics of the output \emph{content} are learned from an image with the desired output content. The resulting convolutional neural networks are merged to yield a network which generates an image with the style of one input image applied to the content of a second input image.

It is clear that style transfer networks are related to the presented work in only the most general way, as both yield outputs with different contrast and same gross structure of input images. It is conceivable that the T$_2$-FLAIR \emph{style} could be learned from a set of T$_2$-FLAIR images from a set of patients and the gross anatomical structure \emph{content} learned from an anatomical image with different, say T$_1$-weighted, contrast from the patient of interest. Importantly, the texture of the output image is based upon the texture profile of the image used to train the \emph{style} portion of the network. As the clinical interpretation of images is often related to the texture of the diagnostic image, this may be suboptimal.

The Build-A-FLAIR network, conversely, approaches the problem of a style transfer in MRI as a non-linear regression model that generates a new contrast image based upon the relative intensities of a set of multi-contrast input images. Thus, while style transfer networks require unique training with each desired \emph{content}, the Build-A-FLAIR network is trained on one set of multi-contrast images and the model is applied to novel images in the synthesis process. This makes the Build-A-FLAIR network dependent upon consistency of input image contrast between the training dataset and the subsequent images on which the neural network is used for inference. Additionally, while style transfer methods work on the scale of a full output image at once, the Build-A-FLAIR network performs inference on an output voxel-by-voxel basis. In doing so, output texture is based only upon the local neighborhoods of a voxel in the input multi-contrast images.

Models including FA as inputs performed worse than models not including FA. While an ideal machine learning algorithm should theoretically yield, at worst, matching results to a model with a subset of inputs as the larger model by giving zero weights to the additional input data, it is apparent that the implemented algorithm reaches a local minimum with non-zero weights. T$_2$-FLAIR contrast, being dependent upon free water and tissue T$_2$ relaxation rate, should not include dependence upon the aniostropy of water diffusion. This result is consistent with the hypothesis that the inclusion of input contrasts with similar physical mechanisms of contrast as the desired output is optimal. 

Qualitatively, the generated T$_2$-FLAIR images exhibit less noise and more smoothness than the acquired T$_2$-FLAIR images. The process of regressing the intensity of each voxel as a function of a series of filters applied to the neighborhood of the voxel is likely responsible for this result. The denoising characteristic of this method is similar to non-local means denoising \citep{nlm} wherein an output voxel is modeled as a weighted combination of regions with similar structure. With a deep convolutional neural network, the fit filters function as the library of similar structures used in non-local means. This neighborhood dependence, as well, is likely responsible for the perceived increase in spatial smoothness.

Recent work has included the generation of synthetic contrast images from individual image contrasts using generalized adversarial networks (SUSAN, \citep{susan}). In that work, contrast changes were meant for data augmentation in machine learning for image segmentation, rather than for the elimination of a given series in a clinical exam. With this different goal, SUSAN is less dependent upon individual image outputs, making compromises from the implementation using a single image contrast as input less detrimental.

A weakness in the development of deep convolutional neural networks is the need for a large volume of training data and significant computational resources for full image training and synthesis. The Build-A-FLAIR network was generated as a local convolutional neural network to perform a regression of a single output voxel value as a function of the neighborhood of voxels in the input multi-contrast images to address both of these challenges. By developing the model as a voxel-wise regression, the hundreds of thousands of voxels in each exam could be used as unique training and validation sets. Further, by modeling input patches rather than full images, the training step can hold more data sets in memory. This allows larger batches in training and improved training convergence, especially when finite memory resources are available \citep{machineLearningTextbook}. The inference problem for each voxel is distinct, allowing the GPU architecture to rapidly perform image synthesis.

With machine learning in medical imaging, the burden of proof to indicate true success that impacts the clinical workflow is high. A concern with contrast generated from artificial intelligence is that the trained model may not represent a pathology which was excluded from training. In this work, a case study of one dataset where a pathology was identified \emph{a priori} is shown wherein the network reproduced the T$_2$-FLAIR white matter hyperintensity even though it was not included in training. This is necessary but not sufficient for translational acceptance. For this method to be implemented in the clinic, a much larger scale prospective study must be performed with a proper blinded radiologist reader scoring. Such work is ongoing.

\begin{figure}[H] 
    \mbox{
    \subfigure[Acquired Axial]{
            \optincludegraphics[width=0.3\textwidth]{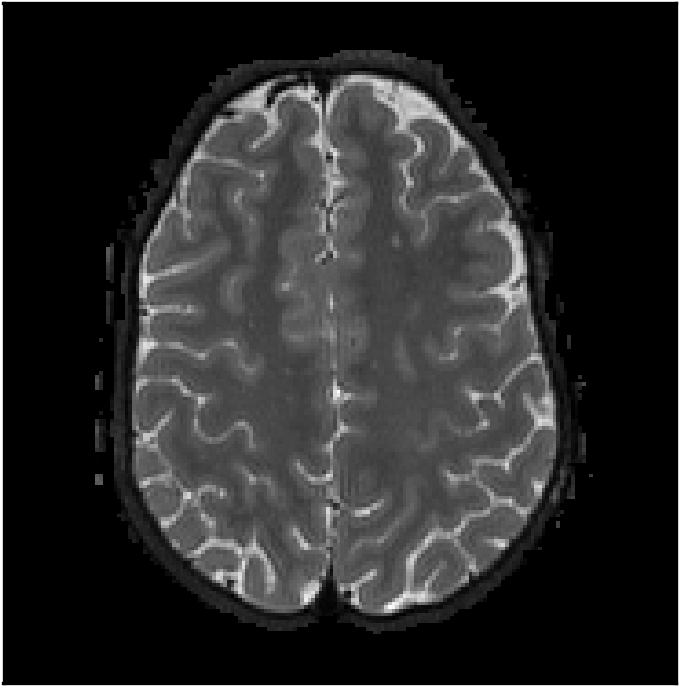}}
    \subfigure[Acquired Coronal]{
        \optincludegraphics[width=0.3\textwidth]{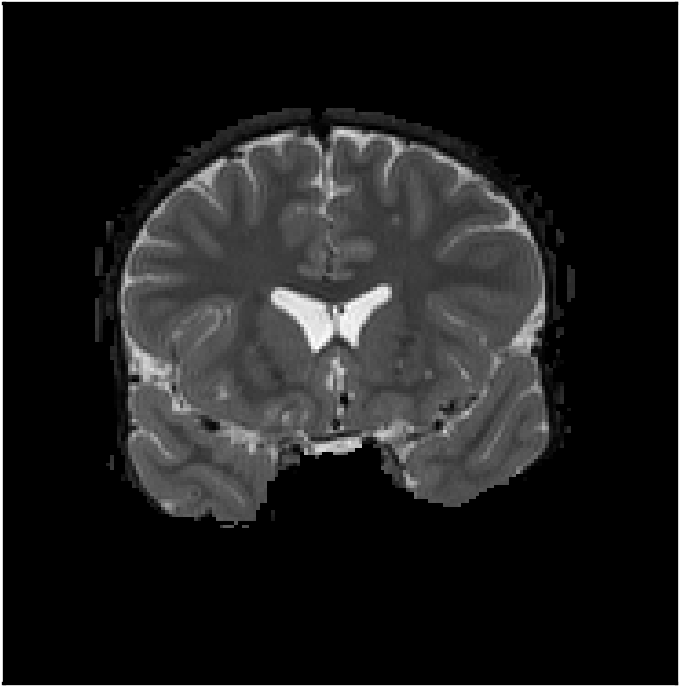}}
    \subfigure[Acquired Sagittal]{
        \optincludegraphics[width=0.3\textwidth]{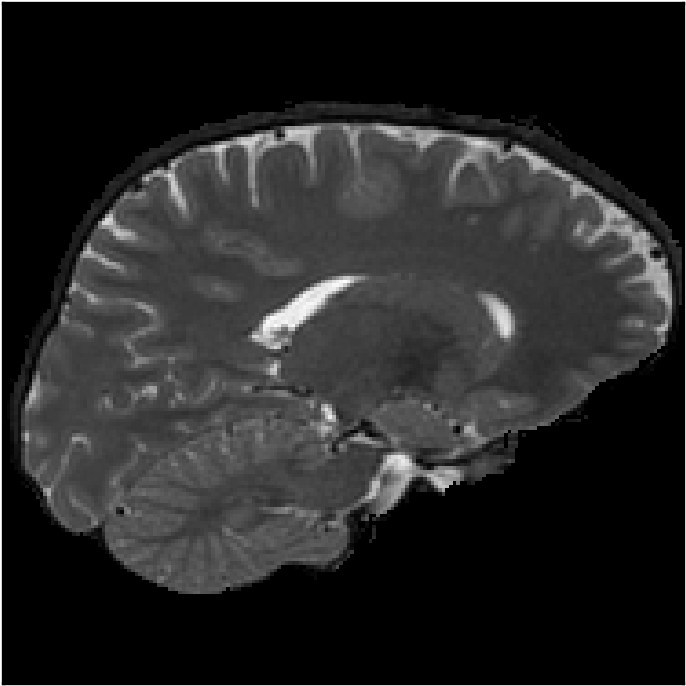}}
    } \\ 
    \mbox{
    \subfigure[Synthetic Axial]{
            \optincludegraphics[width=0.3\textwidth]{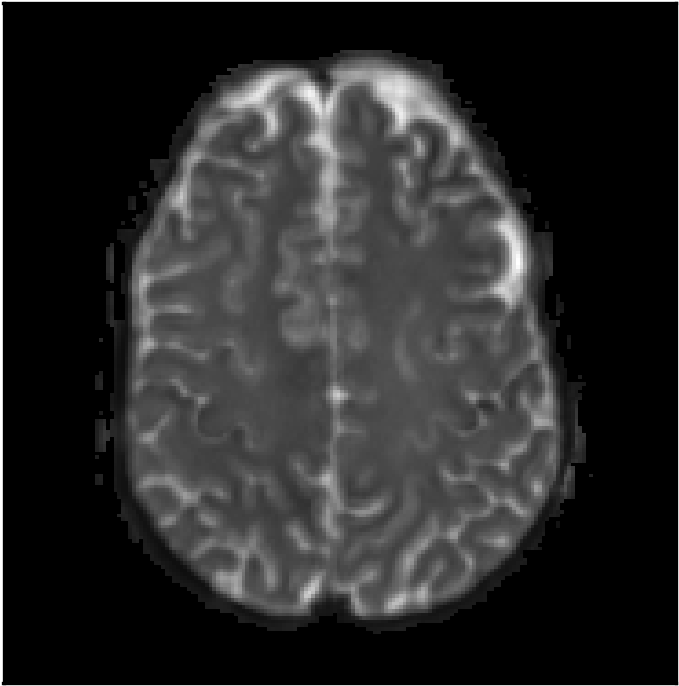}}
    \subfigure[Synthetic Coronal]{
        \optincludegraphics[width=0.3\textwidth]{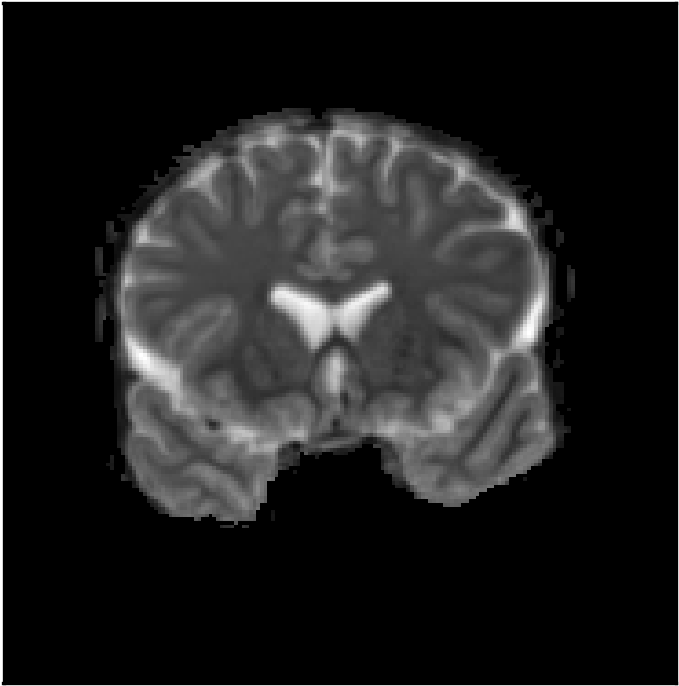}}
    \subfigure[Synthetic Sagittal]{
        \optincludegraphics[width=0.3\textwidth]{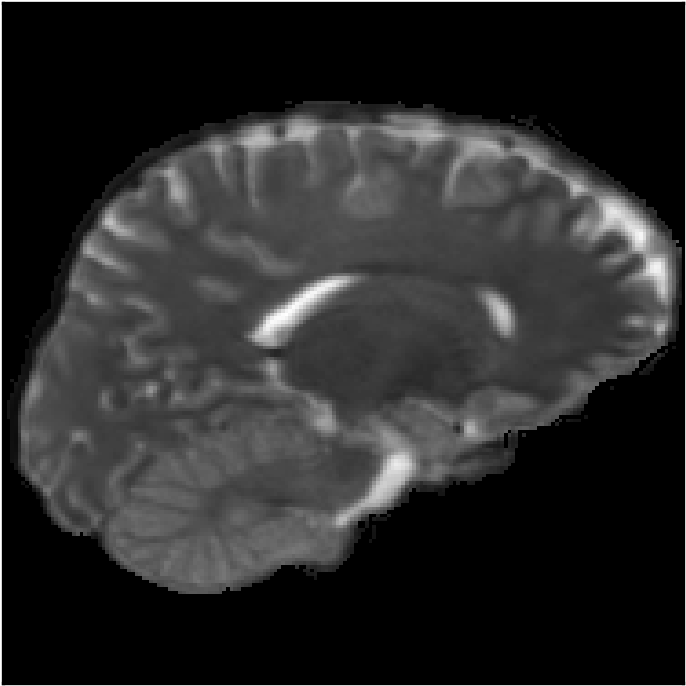}}
    } 
\capt[Synthetic T$_2$-weighted image from re-trained neural network.]{To illustrate the generalizability of the methods presented in this work, a network with an unchanged architecture was trained instead using T$_2$-FLAIR and diffusion-weighted images as inputs to generate a synthetic T$_2$-weighted image. As was shown with synthetic T$_2$-FLAIR generation, input contrasts with related contrasts to the output image were optimal for the model.}
\label{fig:SynthT2}
\end{figure}

While the work herein shows the example of generating synthetic T$_2$-FLAIR images from T$_2$-weighted and diffusion-weighted images, the described network is generalizable. As an example, the network was trained to generate a T$_2$-weighted image from input T$_2$-FLAIR, T$_1$-weighted, and diffusion-weighted images. The result of this model implemented on a dataset not included in the training cohort is shown in Figure \ref{fig:SynthT2}. While the input diffusion-weighted images include a low resolution T$_2$-weighted image, the output synthetic T$_2$-weighted image retains reasonable spatial resolution. As with the generation of T$_2$-FLAIR images, the inclusion of FA as an input does not improve network performance, and the inclusion of a T$_1$-weighted and T$_2$-FLAIR acquisition as input does not drastically improve performance compared to the inclusion of only the T$_2$-FLAIR acquisition. It is expected that other contrasts could be used for training to yield yet other physically related output contrasts. For instance, it is likely that individual echo images from a multiple gradient echo acquisition (like susceptibility weighted imaging) and a T$_2$-weighted spin echo acquisition could yield synthetic T$_2$-weighted Dixon \citep{dixon} fat and water images. Such extensions remain as future potential continuations of this work.

\section{Conclusions}

We demonstrate that T$_2$-FLAIR images can be generated from other standard neuroimaging contrasts.  We further showed that optimal performance was be achieved by the inclusion of contrasts physically related to T$_2$-FLAIR signal in the training dataset. Inputs of T$_2$-weighted and mean diffusivity maps most significantly impacted synthetic T$_2$-FLAIR contrast generation because T$_2$-FLAIR contrast is physically related to both T$_2$ tissue decay rates and the presence of free water.

The presented results are a first step toward the consideration of this synthetically generated contrast to be used to improve MRI value. With T$_2$-FLAIR acquisitions exhibiting reduced signal-to-noise ratio and, thus, requiring increased scan duration, the synthetic generation of T$_2$-FLAIR images from other contrasts could improve MRI value by eliminating the need for a long series in an imaging session. While this proof of concept development is promising, further utility of the Build-A-FLAIR model to replace clinically acquired series requires further, large-scale validation.

\section{Acknowledgments}
This work was supported by the Defense Health Program under the Department of Defense Broad Agency Announcement for Extramural Medical Research through Award No. W81XWH-14-1-0561. Opinions, interpretations, conclusions and recommendations are those of the author and are not necessarily endorsed by the Department of Defense (DHP funds).



\bibliography{references}

\begin{thebibliography}{10}

\bibitem{bernstein}
Bernstein M, King K, Zhou X.
\newblock Handbook of {MRI} Pulse Sequences.
\newblock Elsevier Science, 2004.

\bibitem{bushberg}
Bushberg J.
\newblock The Essential Physics of Medical Imaging.
\newblock Lippincott Williams \& Wilkins, 2002.

\bibitem{ivi}
Feinberg DA, Hoenninger J, Crooks L, Kaufman L, Watts J, Arakawa M.
\newblock Inner Volume {MR} Imaging: Technical Concepts and Their Application.
\newblock Radiology 1985;\hspace{0pt}1563:743--747.

\bibitem{zoom}
Conturo TE, Price RR, Beth AH.
\newblock Rapid Local Rectangular Views and Magnifications: Reduced Phase
  Encoding of Orthogonally Excited Spin Echoes.
\newblock Magnetic Resonance in Medicine 1988;\hspace{0pt}64:418--429.

\bibitem{zoom2}
Wheeler-Kingshott CA, Parker GJ, Symms MR, Hickman SJ, Tofts PS, Miller DH,
  Barker GJ.
\newblock {ADC} Mapping of the Human Optic Nerve: Increased Resolution,
  Coverage, and Reliability with {CSF}-suppressed {ZOOM-EPI}.
\newblock Magnetic Resonance in Medicine 2002;\hspace{0pt}471:24--31.

\bibitem{focus}
Pauly J, Spielman D, Macovski A.
\newblock Echo-Planar Spin-Echo and Inversion Pulses.
\newblock Magnetic Resonance in Medicine 1993;\hspace{0pt}296:776--782.

\bibitem{sense}
Pruessmann KP, Weiger M, Scheidegger MB, Boesiger P.
\newblock {SENSE}: Sensitivity Encoding for Fast {MRI}.
\newblock Magnetic Resonance in Medicine 1999;\hspace{0pt}425:952--962.

\bibitem{grappa}
Griswold MA, Jakob PM, Heidemann RM, Nittka M, Jellus V, Wang J, Kiefer B,
  Haase A.
\newblock Generalized Autocalibrating Partially Parallel Acquisitions
  ({GRAPPA}).
\newblock Magnetic Resonance in Medicine 2002;\hspace{0pt}476:1202--1210.

\bibitem{sms}
Moeller S, Yacoub E, Olman CA, Auerbach E, Strupp J, Harel N, U{\u{g}}urbil K.
\newblock Multiband Multislice {GE-EPI} at 7 Tesla, with 16-Fold Acceleration
  using Partial Parallel Imaging with Application to High Spatial and Temporal
  Whole-Brain {fMRI}.
\newblock Magnetic Resonance in Medicine 2010;\hspace{0pt}635:1144--1153.

\bibitem{cs}
Lustig M, Donoho D, Pauly JM.
\newblock Sparse {MRI}: The Application of Compressed Sensing for Rapid {MR}
  Imaging.
\newblock Magnetic Resonance in Medicine 2007;\hspace{0pt}586:1182--1195.

\bibitem{mlRecon}
Zhu B, Liu JZ, Cauley SF, Rosen BR, Rosen MS.
\newblock Image Reconstruction by Domain-Transform Manifold Learning.
\newblock Nature 2018;\hspace{0pt}5557697:487.

\bibitem{2dTo3d_1}
Gold GE, Busse RF, Beehler C, Han E, Brau AC, Beatty PJ, Beaulieu CF.
\newblock Isotropic {MRI} of the Knee with {3D} Fast Spin-Echo Extended
  Echo-Train Acquisition ({XETA}): Initial Experience.
\newblock American Journal of Roentgenology 2007;\hspace{0pt}1885:1287--1293.

\bibitem{synthMRI1}
Gulani V, Schmitt P, Griswold MA, Webb AG, Jakob PM.
\newblock Towards a Single-Sequence Neurologic Magnetic Resonance Imaging
  Examination: Multiple-Contrast Images from an {IR TrueFISP} Experiment.
\newblock Investigative Radiology 2004;\hspace{0pt}3912:767--774.

\bibitem{synthMRI2}
Warntjes J, Leinhard OD, West J, Lundberg P.
\newblock Rapid Magnetic Resonance Quantification on the Brain: Optimization
  for Clinical Usage.
\newblock Magnetic Resonance in Medicine 2008;\hspace{0pt}602:320--329.

\bibitem{fingerprinting}
Ma D, Gulani V, Seiberlich N, Liu K, Sunshine JL, Duerk JL, Griswold MA.
\newblock Magnetic Resonance Fingerprinting.
\newblock Nature 2013;\hspace{0pt}4957440:187.

\bibitem{h2h2_1}
Kaushal M, Espa{\~n}a LY, Nencka AS, Wang Y, Nelson LD, McCrea MA, Meier TB.
\newblock Resting-State Functional Connectivity After Concussion is Associated
  with Clinical Recovery.
\newblock Human Brain Mapping 2018;\hspace{0pt}.

\bibitem{h2h2_2}
Nelson LD, Kramer MD, Patrick CJ, McCrea MA.
\newblock Modeling the Structure of Acute Sport-Related Concussion Symptoms: A
  Bifactor Approach.
\newblock Journal of the International Neuropsychological Society
  2018;\hspace{0pt}248:793--804.

\bibitem{h2h2_3}
Meier TB, Nelson LD, Huber DL, Bazarian JJ, Hayes RL, McCrea MA.
\newblock Prospective Assessment of Acute Blood Markers of Brain Injury in
  Sport-Related Concussion.
\newblock Journal of neurotrauma 2017;\hspace{0pt}3422:3134--3142.

\bibitem{dcm2nii}
Li X, Morgan PS, Ashburner J, Smith J, Rorden C.
\newblock The First Step for Neuroimaging Data Analysis: {DICOM} to {NIfTI}
  Conversion.
\newblock Journal of Neuroscience Methods 2016;\hspace{0pt}264:47--56.

\bibitem{fsl}
Jenkinson M, Beckmann CF, Behrens TE, Woolrich MW, Smith SM.
\newblock {FSL}.
\newblock {NeuroImage} 2012;\hspace{0pt}622:782--790.

\bibitem{flirt}
Jenkinson M, Smith S.
\newblock A Global Optimisation Method for Robust Affine Registration of Brain
  Images.
\newblock Medical Image Analysis 2001;\hspace{0pt}52:143--156.

\bibitem{python}
van Rossum G, de~Boer J.
\newblock Interactively Testing Remote Servers Using the Python Programming
  Language.
\newblock {CWI} Quarterly 1991;\hspace{0pt}44:283--303.

\bibitem{numpy}
Walt Svd, Colbert SC, Varoquaux G.
\newblock The {NumPy} Array: A Structure for Efficient Numerical Computation.
\newblock Computing in Science \& Engineering 2011;\hspace{0pt}132:22--30.

\bibitem{nibabel}
Brett M, Hanke M, Cipollini B, C{\^o}t{\'e} MA, Markiewicz C, Gerhard S, Larson
  E, Lee GR, Halchenko Y, Kastman E, {\em et~al.\/}.
\newblock NIBabel: 2.1. 0.
\newblock Zenodo 2016;\hspace{0pt}.

\bibitem{pytorch}
Paszke A, Gross S, Chintala S, Chanan G, Yang E, DeVito Z, Lin Z, Desmaison A,
  Antiga L, Lerer A.
\newblock Automatic Differentiation in Pytorch 2017;\hspace{0pt}.

\bibitem{relu}
Nair V, Hinton GE.
\newblock Rectified Linear Units Improve Restricted Boltzmann Machines.
\newblock Proceedings of the 27th International Conference on Machine Learning
  (ICML-10).
\newblock 2010 807--814.

\bibitem{dropout}
Srivastava N, Hinton G, Krizhevsky A, Sutskever I, Salakhutdinov R.
\newblock Dropout: A Simple Way to Prevent Neural Networks from Overfitting.
\newblock The Journal of Machine Learning Research
  2014;\hspace{0pt}151:1929--1958.

\bibitem{adam}
Kingma DP, Ba J.
\newblock Adam: A Method for Stochastic Optimization.
\newblock arXiv preprint arXiv:14126980 2014;\hspace{0pt}.

\bibitem{ssim}
Wang Z, Bovik AC, Sheikh HR, Simoncelli EP.
\newblock Image Quality Assessment: From Error Visibility to Structural
  Similarity.
\newblock {IEEE} Transactions on Image Processing
  2004;\hspace{0pt}134:600--612.

\bibitem{styleTransfer}
Gatys LA, Ecker AS, Bethge M.
\newblock Image Style Transfer Using Convolutional Neural Networks.
\newblock The {IEEE} Conference on Computer Vision and Pattern Recognition
  ({CVPR}).
\newblock 2016 .

\bibitem{nlm}
Buades A, Coll B, Morel JM.
\newblock A Non-Local Algorithm for Image Denoising.
\newblock IEEE Computer Society Conference on Computer Vision and Pattern
  Recognition, 2005. CVPR 2005.
\newblock 2005 60--65.

\bibitem{susan}
Liu F.
\newblock SUSAN: Segment Unannotated Image Structure using Adversarial Network.
\newblock Magnetic Resonance in Medicine ;\hspace{0pt}00.

\bibitem{machineLearningTextbook}
G{\'e}ron A.
\newblock Hands-on Machine Learning with Scikit-Learn and TensorFlow: Concepts,
  Tools, and Techniques to Build Intelligent Systems.
\newblock O'Reilly Media, Inc., 2017.

\bibitem{dixon}
Dixon WT.
\newblock Simple Proton Spectroscopic Imaging.
\newblock Radiology 1984;\hspace{0pt}1531:189--194.

\end{thebibliography}

\section{Figures and Tables}

\begin{table}[H] 
\hspace{-0.3in}
\begin{tabular}{|C{1.cm} |C{2.5cm}|C{2.5cm} |C{2.5cm} |C{2.5cm} |C{2.5cm} || c|}
\hline
Model & T$_2$-Weighted 		& T$_1$-Weighted 		& MD 			& FA 			& S0 			& SSIM \\
\hline
1 & \cellcolor{white}Included 	& \cellcolor{gray}Omitted		& \cellcolor{gray}Omitted		& \cellcolor{gray}Omitted		& \cellcolor{gray}Omitted		& 0.69358 \\
\hline
2 & \cellcolor{gray}Omitted 	& \cellcolor{white}Included		& \cellcolor{gray}Omitted		& \cellcolor{gray}Omitted		& \cellcolor{gray}Omitted		& 0.76663 \\
\hline
3 & \cellcolor{gray}Omitted 	& \cellcolor{gray}Omitted		& \cellcolor{white}Included		& \cellcolor{white}Included		& \cellcolor{white}Included		& 0.81578 \\
\hline
4 & \cellcolor{white}Included 	& \cellcolor{white}Included		& \cellcolor{gray}Omitted		& \cellcolor{gray}Omitted		& \cellcolor{gray}Omitted		& 0.81756 \\
\hline
5 & \cellcolor{gray}Omitted 	& \cellcolor{white}Included		& \cellcolor{white}Included		& \cellcolor{white}Included		& \cellcolor{white}Included		& 0.87856 \\
\hline
6 & \cellcolor{gray}Omitted 	& \cellcolor{white}Included		& \cellcolor{white}Included		& \cellcolor{gray}Omitted		& \cellcolor{white}Included		& 0.88312 \\
\hline
7 & \cellcolor{white}Included 	& \cellcolor{gray}Omitted		& \cellcolor{white}Included		& \cellcolor{white}Included		& \cellcolor{white}Included		& 0.89064 \\
\hline
8 & \cellcolor{white}Included 	& \cellcolor{gray}Omitted		& \cellcolor{white}Included		& \cellcolor{gray}Omitted		& \cellcolor{white}Included		& 0.90043 \\
\hline
9 & \cellcolor{white}Included 	& \cellcolor{white}Included		& \cellcolor{white}Included		& \cellcolor{white}Included		& \cellcolor{white}Included		& 0.90584 \\
\hline
10 & \cellcolor{white}Included 	& \cellcolor{white}Included		& \cellcolor{white}Included		& \cellcolor{gray}Omitted		& \cellcolor{white}Included		& 0.90881 \\
\hline
\end{tabular}

\capt[Models evaluated in the Build-A-FLAIR framework.]{Ten separate models were developed with varying inputs to generate synthetic T$_2$-FLAIR contrast. The models included subsets of input contrasts including T$_2$-weighted, T$_1$-weighted, mean diffusivity (MD), fractional anisotropy (FA), and non-diffusion weighted images (S0). Each row of this table represents one model tested in this work. Contrasts used in each model are shown in white in the line for the model, and contrasts not included as inputs are shown as gray. Models are listed in order of increasing performance, as measured by the structural similarity index between the synthetic T$_2$-FLAIR image and the acquired T$_2$-FLAIR image in the subject which was fully removed from the training process (V1).}
\label{tab:models}
\end{table}


\renewcommand{\thefigure}{S\arabic{figure}}
\def\table{\def\figurename{Table}\figure}
\setcounter{figure}{0}




\end{document}